\documentclass[twoside,english,a4paper]{article}
\usepackage[T1]{fontenc}
\usepackage[latin9]{inputenc}
\usepackage{array}
\usepackage{float}
\usepackage{amsmath}
\usepackage{graphicx}

\makeatletter

\providecommand{\tabularnewline}{\\}
\floatstyle{ruled}
\newfloat{algorithm}{tbp}{loa}
\providecommand{\algorithmname}{Algorithm}
\floatname{algorithm}{\protect\algorithmname}

\renewcommand{\textemdash}{---}

\@ifundefined{showcaptionsetup}{}{%
 \PassOptionsToPackage{caption=false}{subfig}}
\usepackage{subfig}
\makeatother

\usepackage{babel}
\begin{document}

\title{$G^{1}$ hole filling with S-patches made easy}


\author{Péter Salvi\\Budapest University of Technology and Economics}


\maketitle
\begin{abstract}
S-patches have been around for 30 years, but they are seldom used,
and are considered more of a mathematical curiosity than a practical
surface representation. In this article a method is presented for
automatically creating S-patches of any degree or any number of sides,
suitable for inclusion in a curve network with tangential continuity
to the adjacent surfaces. The presentation aims at making the implementation
straightforward; a few examples conclude the paper.
\end{abstract}

\section{Introduction}

\begin{figure}
\begin{centering}
\includegraphics[width=0.7\textwidth]{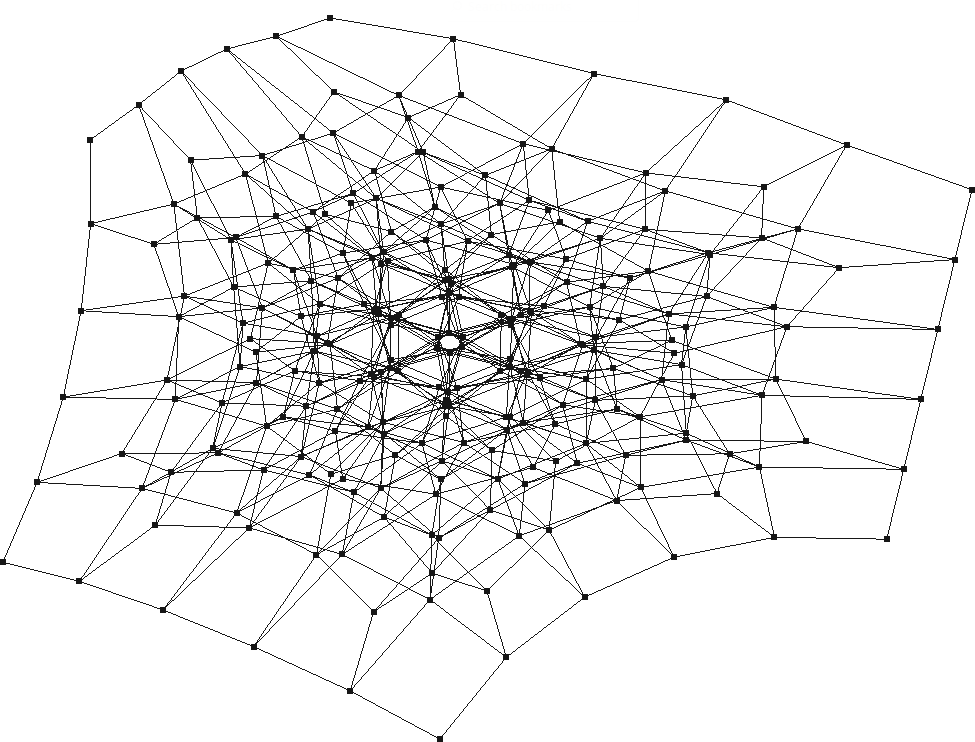}
\par\end{centering}
\caption{\label{fig:6-5-cnet}The control net of a 6-sided quintic S-patch.}
\end{figure}

The S-patch~\cite{Loop:1989} is the natural multi-sided generalization
of the Bézier triangle. It has various nice mathematical properties,
and has been known for a long time, but it has not become a standard
surface representation like the tensor product Bézier patch. It is
easy to see why: the number of control points grows very fast, and
the control network becomes intricate even for surfaces of moderate
complexity. For example, a 6-sided quintic patch has 252 control points
(see Figure~\ref{fig:6-5-cnet}), while a tensor product quintic
surface has only 36.

Interactive placement of such a large number of points is hardly possible,
but this does not mean that the whole representation should be discarded
as impractical. As we will see in the rest of the paper, it is particularly
suitable for filling multi-sided holes while ensuring tangential continuity
to adjacent surfaces\textemdash an important problem in curve network
based design.

\section{Previous work}

A solution for hole filling with S-patches was first presented in~\cite{Loop:1990},
where quadratic and cubic surfaces are created based on Sabin nets.
Although the equations shown in the paper are also valid for higher
degrees, some of the details are left for the reader to figure out.
The placement of interior control points, in particular, is not discussed
for degrees above cubic.

A recent S-patch-based construction~\cite{Hettinga:2018} gives a
simple mechanism for $G^{1}$-con\-ti\-nu\-o\-us hole filling,
even without twist constraints, but only for cubic boundaries. Also,
there are no details on the placement of interior control points.

Generalized Bézier (GB) patches~\cite{Varady:2016:EG} present another
approach to represent multi-sided Bézier surfaces. Tangential connection
to adjacent patches is easily achieved, and its simple control point
structure makes it possible to edit the surface interior interactively.
On the other hand, some common operations, like analytic derivative
computation or exact degree elevation, are difficult or not even possible.

Transfinite surfaces (e.g.~\cite{Salvi:2017:WAIT}) can also be used
to fill multi-sided holes when the boundaries are not necessarily
Bézier curves; some even have a limited control over the interior.

\section{\label{sec:S-patches}S-patches}

An $n$-sided S-patch is defined over an $n$-gon, parameterized by
generalized ba\-ry\-cen\-tric coordinates $\boldsymbol{\lambda}=(\lambda_{1},\dots,\lambda_{n})$,
e.g.~Wachspress or mean value coordinates. (In the current paper,
we will always use a regular polygon as the domain.) Its control points
$\{P_{\mathbf{s}}\}$ are labeled by $n$ non-negative integers whose
sum is the degree\footnote{Also called the \emph{depth} of the S-patch, to differentiate it from
the (rational) polynomial degree of the surface. We will not make
this distinction here.} of the surface ($d$). We will use the notation $L_{n,d}$ for the
set of all such labels, and $(s_{1},\dots,s_{n})=\mathbf{s}\in L_{n,d}$
for one particular label. It is easy to see that $|L_{n,d}|=\left(\!\!{n \choose d}\!\!\right)=\binom{n+d-1}{d}$.

The surface point corresponding to a domain point with barycentric
coordinates $\boldsymbol{\lambda}$ is defined as
\begin{equation}
S(\boldsymbol{\lambda})=\sum_{\mathbf{s}\in L_{n,d}}P_{\mathbf{s}}\cdot B_{\mathbf{s}}^{d}(\boldsymbol{\lambda})=\sum_{\mathbf{s}\in L_{n,d}}P_{\mathbf{s}}\cdot\frac{d!}{\prod_{i=1}^{n}s_{i}!}\cdot\prod_{i=1}^{n}\lambda_{i}^{s_{i}},\label{eq:spatch}
\end{equation}
where $B_{\mathbf{s}}^{d}(\boldsymbol{\lambda})$ are Bernstein polynomials
with multinomial coefficients.

The labeling system deserves closer inspection. We define the shift
operation $\sigma_{j}^{\pm}(\mathbf{s})$ as decrementing $s_{j}$
and incrementing $s_{j\pm1}$ in label $\mathbf{s}$, e.g.~$\sigma_{2}^{+}(1,1,1)=(1,0,2)$
and $\sigma_{1}^{-}(1,0,2)=(0,0,3)$. (Note that label indexing is
cyclical.) $P_{\mathbf{s}}$ is said to be \emph{adjacent} or \emph{connected}
to $P_{\mathbf{\hat{s}}}$, if $\sigma_{j}^{\pm}(\mathbf{s})=\mathbf{\hat{s}}$
for some $j$.

\begin{figure}
\begin{centering}
\includegraphics[width=0.8\textwidth]{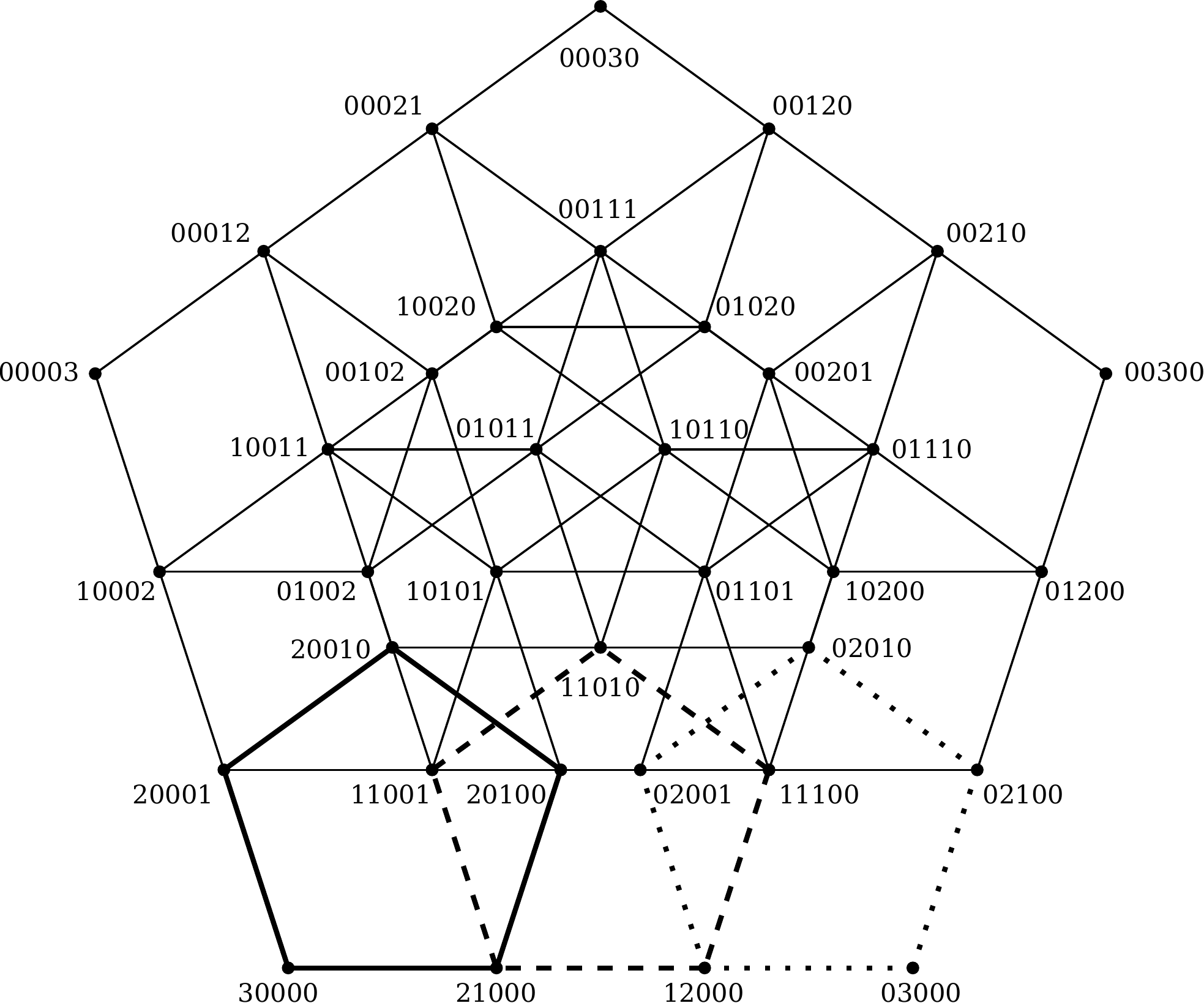}
\par\end{centering}
\caption{\label{fig:5-3-cnet}The control net of a 5-sided cubic S-patch with
labels. The boundary panels on side 1 are highlighted.}
\end{figure}

Control points with labels such that $s_{i}+s_{i+1}=d$ are on the
$i$-th boundary, e.g.~$(0,0,2,3,0,0)\in I_{6,5}$ or $(2,0,0,0,1)\in I_{5,3}$.
Specifically, the $j$-th point on the boundary has a label for which
$s_{i}=d-j$ and $s_{i+1}=j$, $0\leq j\leq d$ (see Figure~\ref{fig:5-3-cnet}).
We will also refer to these labels by the notation $\mathbf{s}_{i,j}$.

Differentiation of Eq.~(\ref{eq:spatch}) shows that the first cross
derivative at a boundary point is a combination of control points
with labels such that $s_{i}+s_{i+1}\geq d-1$ for some $i$. These
can be grouped on each side into $d$ \emph{boundary panels}. The
$j$-th panel consists of $n$ points: 
\[
\mathbf{s}_{i,j},\ \sigma_{j}^{+}(\mathbf{s}_{i,j}),\ \sigma_{j+1}^{+}(\sigma_{j}^{+}(\mathbf{s}_{i,j})),\ \sigma_{j+2}^{+}(\sigma_{j+1}^{+}(\sigma_{j}^{+}(\mathbf{s}_{i,j}))),\ \dots,
\]
until we get back to $\mathbf{s}_{i,j}$. The $k$-th point of this
panel is also denoted by $P_{j,k}^{i}$, so $P_{j,2}^{i}=P_{j+1,1}^{i}$.
For example, in the five-sided cubic patch in Figure~\ref{fig:5-3-cnet},
$\{P_{2,k}^{3}\}$ contains control points with labels $(0,0,1,2,0)$,
$(0,0,0,3,0)$, $(0,0,0,2,1)$, $(1,0,0,2,0)$, and $(0,1,0,2,0)$.
\begin{figure}
\begin{centering}
\subfloat[\label{fig:Boundary-constraints}Boundary constraints]{\begin{centering}
\includegraphics[height=0.28\textheight]{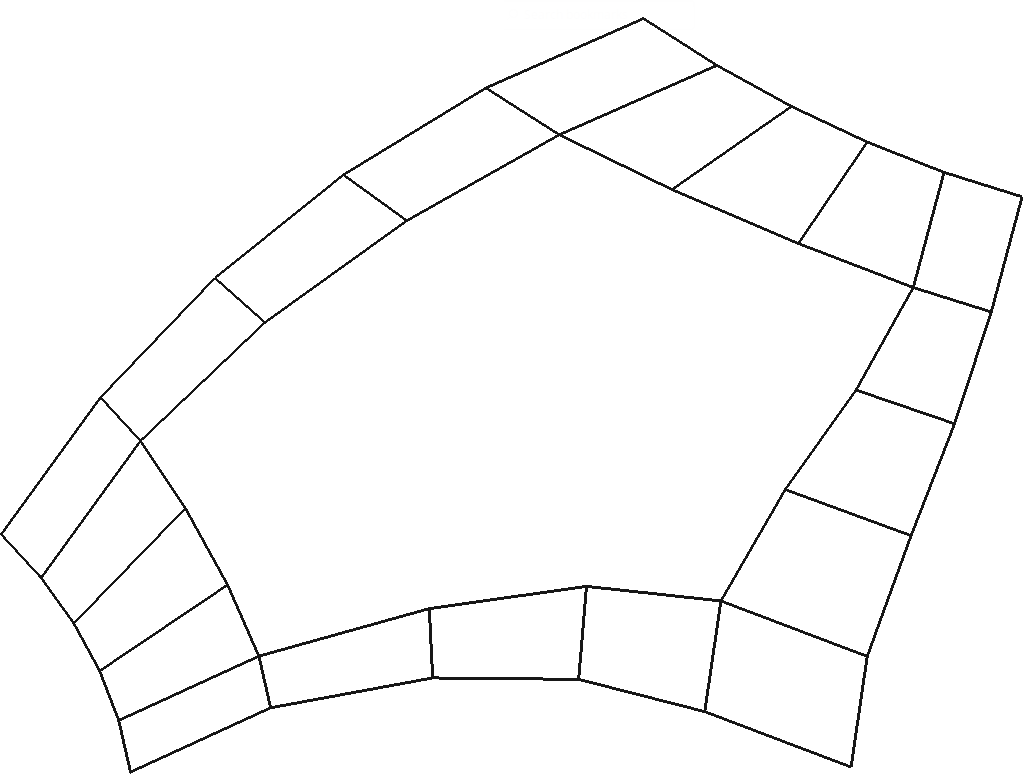}
\par\end{centering}
}
\par\end{centering}
\begin{centering}
\subfloat[\label{fig:G1-boundary-panels}$G^{1}$ boundary panels]{\begin{centering}
\includegraphics[height=0.28\textheight]{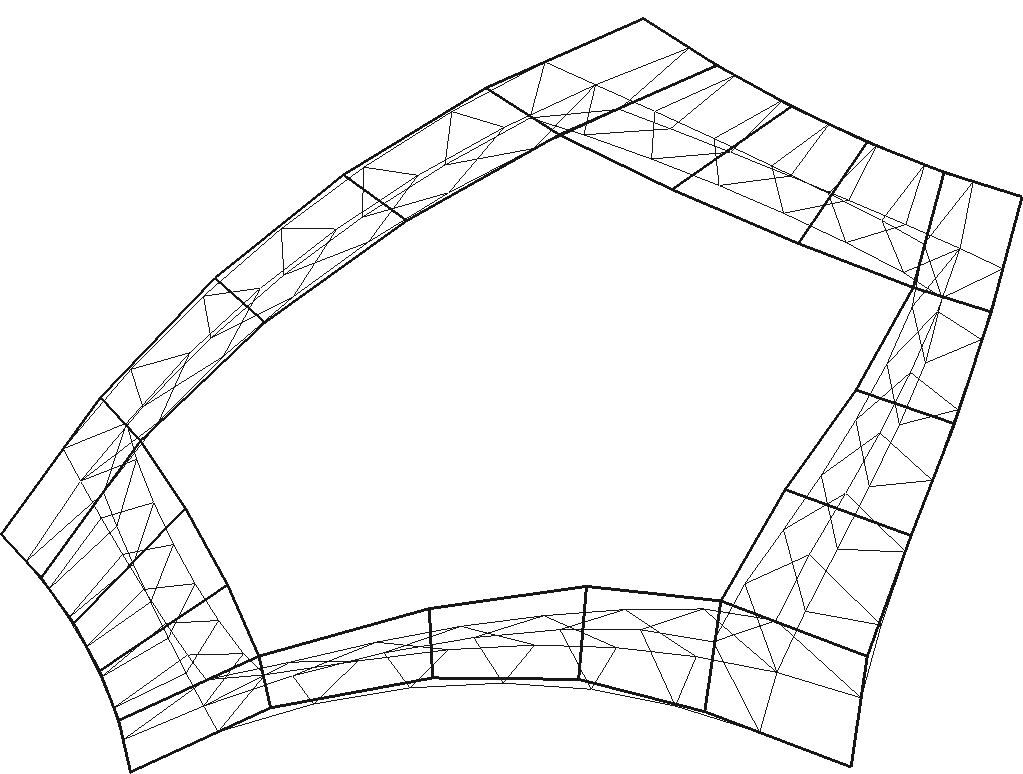}
\par\end{centering}
}
\par\end{centering}
\begin{centering}
\subfloat[\label{fig:Interior-control-points}Interior control points]{\begin{centering}
\includegraphics[height=0.28\textheight]{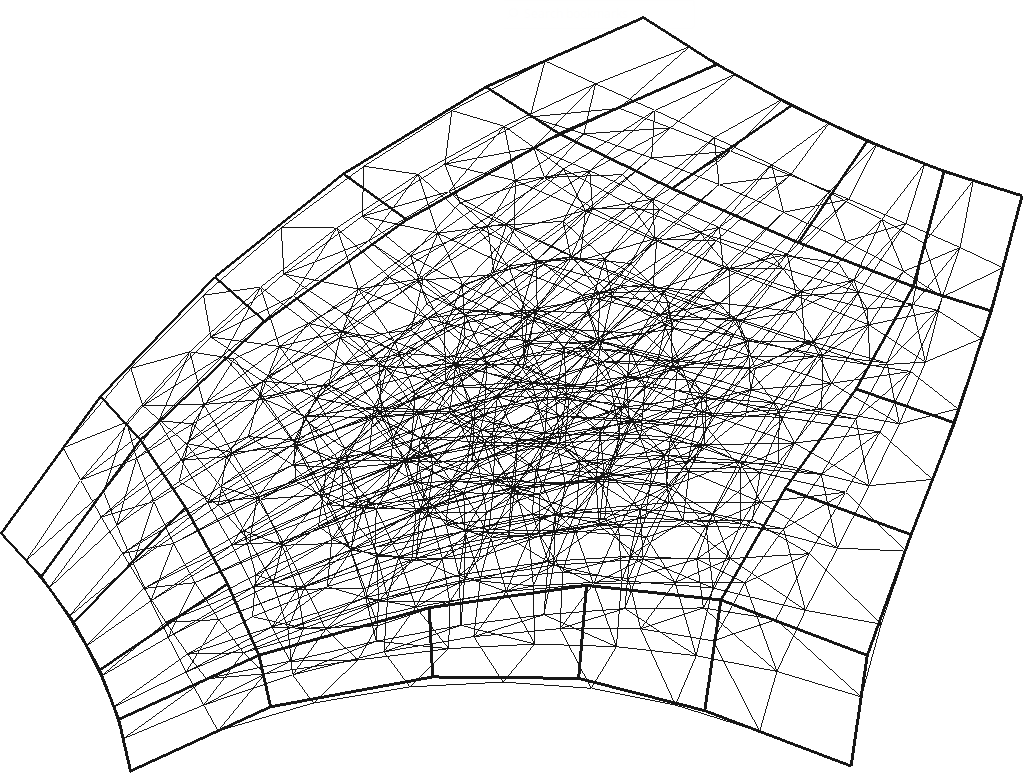}
\par\end{centering}
}
\par\end{centering}
\caption{Stages of filling a five-sided hole with quintic boundary constraints.}
\end{figure}

\section{Hole filling}

Now that S-patches are defined, we can turn our attention to the $n$-sided
hole filling problem. Let the boundary curves be given in Bézier form,
with a common degree $d$. Cross-derivative constraints are represented
by a second row of control points on all sides; the two control rows
together are called a \emph{Bézier ribbon}~\cite{Varady:2016:EG}.
We need to construct a surface that matches, for each side, the derivatives
of a $d\times d$-degree tensor product Bézier patch with the same
two control rows, see Figure~\ref{fig:Boundary-constraints}.

Let $C_{j,k}^{i}$ denote the $j$-th control point in the $k$-th
row on the $i$-th side ($1\leq i\leq n$, $0\leq j\leq d$, $0\leq k\leq1$).
We assume that the constraints are twist-compatible, i.e., $C_{1,1}^{i}=C_{d-1,1}^{i-1}$,
$C_{1,0}^{i}=C_{d,1}^{i-1}$ and $C_{0,1}^{i}=C_{d-1,0}^{i-1}$ for
each $i$ (using cyclic indexing). This form of the boundary constraints
is called a \emph{Sabin net} in~\cite{Loop:1990}.

The rest of this section will describe how to compute the correct
boundary panels for $G^{1}$ interpolation of the cross-derivatives
(Figure~\ref{fig:G1-boundary-panels}), and how to set the position
of the remaining control points automatically (Figure~\ref{fig:Interior-control-points}).

\subsection{$G^{1}$ boundaries}

This section is based on~\cite{Loop:1990}, where the derivation
of these equations can be found. In the paper explicit formulas are
only given for the special cases of $d=2$ and~$3$; here we will
look at the general degree case for the convenience of the reader.

Before diving into the intricacies of tangential continuity, it is
important to note that the boundaries of an S-patch are Bézier curves
with $P_{j,1}^{i}$ ($0\leq j\leq d$) as the control polygons, so
if positional ($C^{0}$) interpolation is enough, it can be done by
setting $P_{j,1}^{i}=C_{j,0}^{i}$ for every $i$ and $j$.

The conditions of tangential continuity with a Bézier triangle (or
another S-patch) of the same degree are simple: (i) each boundary
panel should be the affine image of the domain polygon, and (ii) opposite
boundary panels along the common boundary should be in the same plane.

From (i) it follows that it suffices to set 3 points of each panel,
2 of which are on the boundary, so only one extra point is needed,
e.g.~$P_{j,n}^{i}$. The $n$-sided setting introduces additional
constraints, which can be satisfied only with an S-patch of degree
$d+3$. Eventually we arrive at the equation
\begin{align*}
P_{j,n}^{i} & =P_{j,1}^{i}+\frac{d}{(d+3)}\left[\delta_{1}^{d}(j)\left(C_{j,0}^{i}-C_{j-1,0}^{i}\right)\cdot2c\binom{d-1}{j-1}+\right.\\
 & \qquad\qquad\qquad\qquad\delta_{2}^{d+1}(j)\left(C_{j-1,0}^{i}-C_{j-2,0}^{i}\right)\cdot4c\binom{d-1}{j-2}+\\
 & \qquad\qquad\qquad\qquad\delta_{3}^{d+2}(j)\left(C_{j-2,0}^{i}-C_{j-3,0}^{i}\right)\cdot2c\binom{d-1}{j-3}+\\
 & \qquad\qquad\qquad\qquad\delta_{0}^{d}(j)\left(C_{j,1}^{i}-C_{j,0}^{i}\right)\cdot\binom{d}{j}+\\
 & \qquad\qquad\qquad\qquad\delta_{1}^{d+1}(j)\left(C_{j-1,1}^{i}-C_{j-1,0}^{i}\right)\cdot(2+2c)\binom{d}{j-1}+
\end{align*}
\begin{equation}
\left.\qquad\qquad\qquad\qquad\delta_{2}^{d+2}(j)\left(C_{j-2,1}^{i}-C_{j-2,0}^{i}\right)\cdot\binom{d}{j-2}\right]/\binom{d+2}{j},
\end{equation}
where $c=-\cos(2\pi/n)$ and 
\begin{equation}
\delta_{l}^{h}(j)=\begin{cases}
1 & l\leq j\leq h\\
0 & \text{otherwise}
\end{cases}
\end{equation}
(The boundary control points $P_{j,1}^{i}$ can be computed as in
the $C^{0}$ case, after degree elevating the Bézier ribbon 3 times.)

The remaining control points in a panel are computed by the affine
transformation. After solving
\begin{equation}
\left[\begin{array}{ccc}
1 & 0 & 1\\
\cos(2\pi/n) & \sin(2\pi/n) & 1\\
\cos(4\pi/n) & \sin(4\pi/n) & 1
\end{array}\right]\cdot M^{T}=\left[\begin{array}{c}
(P_{j,n}^{i})^{T}\\
(P_{j,1}^{i})^{T}\\
(P_{j,2}^{i})^{T}
\end{array}\right]
\end{equation}
for $M^{T}$, we get
\begin{equation}
P_{j,k}^{i}=M\cdot\left[\begin{array}{ccc}
\cos(2k\pi/n) & \sin(2k\pi/n) & 1\end{array}\right]^{T}.
\end{equation}

\subsection{Interior control points}

There are various heuristics for the placement of interior control
points, mainly for lower-degree patches. The procedure presented here
works for any S-patch, independently of the number of sides, the degree,
or the type of continuity. It is based on the work of Farin et al.~\cite{Farin:1999},
which shows that applying discrete masks on the control network of
a tensor product Bézier surface can result, depending on the masks
used, in a Coons patch or a quasi-minimal surface. Monterde et al.~\cite{Monterde:2004}
investigated harmonic and biharmonic masks, and these are easily applicable
to S-patches, as well.

Table~\ref{tab:Harmonic-mask} shows the harmonic mask. For a tensor
product Bézier patch with control points $Q_{i,j}$, this represents
the equation system 
\begin{equation}
Q_{i+1,j}+Q_{i-1,j}+Q_{i,j+1}+Q_{i,j-1}-4Q_{i,j}=0,
\end{equation}
i.e., each control point should be the average of its neighbors. If
we fix the outermost control rows, the system can be solved, resulting
in a smooth distribution of points. This can be directly applied to
S-patches, using the adjacency relation defined in Section~\ref{sec:S-patches}.
Note, however, that there can be different masks even within one patch,
as the valency of control points may vary.

\begin{table}
\hfill{}\subfloat[\label{tab:Harmonic-mask}Harmonic mask]{\begin{centering}
\begin{minipage}[b][1\totalheight][t]{0.4\textwidth}%
\begin{center}
\begin{tabular}{>{\centering}p{2em}|>{\centering}p{2em}|>{\centering}p{2em}|>{\centering}p{2em}|>{\centering}p{2em}}
\multicolumn{1}{>{\centering}p{2em}}{} & \multicolumn{1}{>{\centering}p{2em}}{} & \multicolumn{1}{>{\centering}p{2em}}{} & \multicolumn{1}{>{\centering}p{2em}}{} & \tabularnewline
\cline{2-4} 
 & 0 & 1 & 0 & \tabularnewline
\cline{2-4} 
 & 1 & \textbf{-4} & 1 & \tabularnewline
\cline{2-4} 
 & 0 & 1 & 0 & \tabularnewline
\cline{2-4} 
\multicolumn{1}{>{\centering}p{2em}}{} & \multicolumn{1}{>{\centering}p{2em}}{} & \multicolumn{1}{>{\centering}p{2em}}{} & \multicolumn{1}{>{\centering}p{2em}}{} & \tabularnewline
\end{tabular}
\par\end{center}%
\end{minipage}
\par\end{centering}
}\hfill{}\subfloat[\label{tab:Biharmonic-mask}Biharmonic mask]{\begin{centering}
\begin{minipage}[b][1\totalheight][t]{0.4\textwidth}%
\begin{center}
\begin{tabular}{|>{\centering}p{2em}|>{\centering}p{2em}|>{\centering}p{2em}|>{\centering}p{2em}|>{\centering}p{2em}|}
\hline 
0 & 0 & 1 & 0 & 0\tabularnewline
\hline 
0 & 2 & -8 & 2 & 0\tabularnewline
\hline 
1 & -8 & \textbf{20} & -8 & 1\tabularnewline
\hline 
0 & 2 & -8 & 2 & 0\tabularnewline
\hline 
0 & 0 & 1 & 0 & 0\tabularnewline
\hline 
\end{tabular}
\par\end{center}%
\end{minipage}
\par\end{centering}
}\hfill{}

\caption{Masks used on tensor product Bézier control networks. The number in
bold is the coefficient of the control point the mask is applied to.}
\end{table}
When the cross-derivatives are fixed at the boundary, it is natural
to use a biharmonic mask (see Table~\ref{tab:Biharmonic-mask}).
This is computed by applying the harmonic mask on itself, which translates
to S-patches in a straightforward way, see Algorithm~\ref{alg:Computing-masks}.
All examples in this paper were computed by this method.
\begin{algorithm}
\begin{minipage}[t]{0.5\columnwidth}%
\texttt{harmonicMask(i):}

\texttt{\phantom{\texttt{xx}}neighbors = }indices adjacent to\texttt{
i}

\texttt{\phantom{\texttt{xx}}for j in neighbors:}

\texttt{\phantom{\texttt{xx}}\phantom{\texttt{xx}}mask{[}j{]} =
1}

\texttt{\phantom{\texttt{xx}}mask{[}i{]} = -length(neighbors)}

\texttt{\phantom{\texttt{xx}}return mask}%
\end{minipage}%
\begin{minipage}[t]{0.5\columnwidth}%
\texttt{biharmonicMask(i):}

\texttt{\phantom{\texttt{xx}}mask = 0 }for all indices

\texttt{\phantom{\texttt{xx}}for (j,wj) in harmonicMask(i):}

\texttt{\phantom{\texttt{xx}}\phantom{\texttt{xx}}for (k,wk) in
harmonicMask(j):}

\texttt{\phantom{\texttt{xx}}\phantom{\texttt{xx}}\phantom{\texttt{xx}}mask{[}k{]}
+= wj {*} wk}

\texttt{\phantom{\texttt{xx}}return mask}%
\end{minipage}

\caption{\label{alg:Computing-masks}Computing harmonic and biharmonic masks.}

\end{algorithm}

\section{Examples}

\begin{figure}
\begin{centering}
\includegraphics[width=0.63\textwidth]{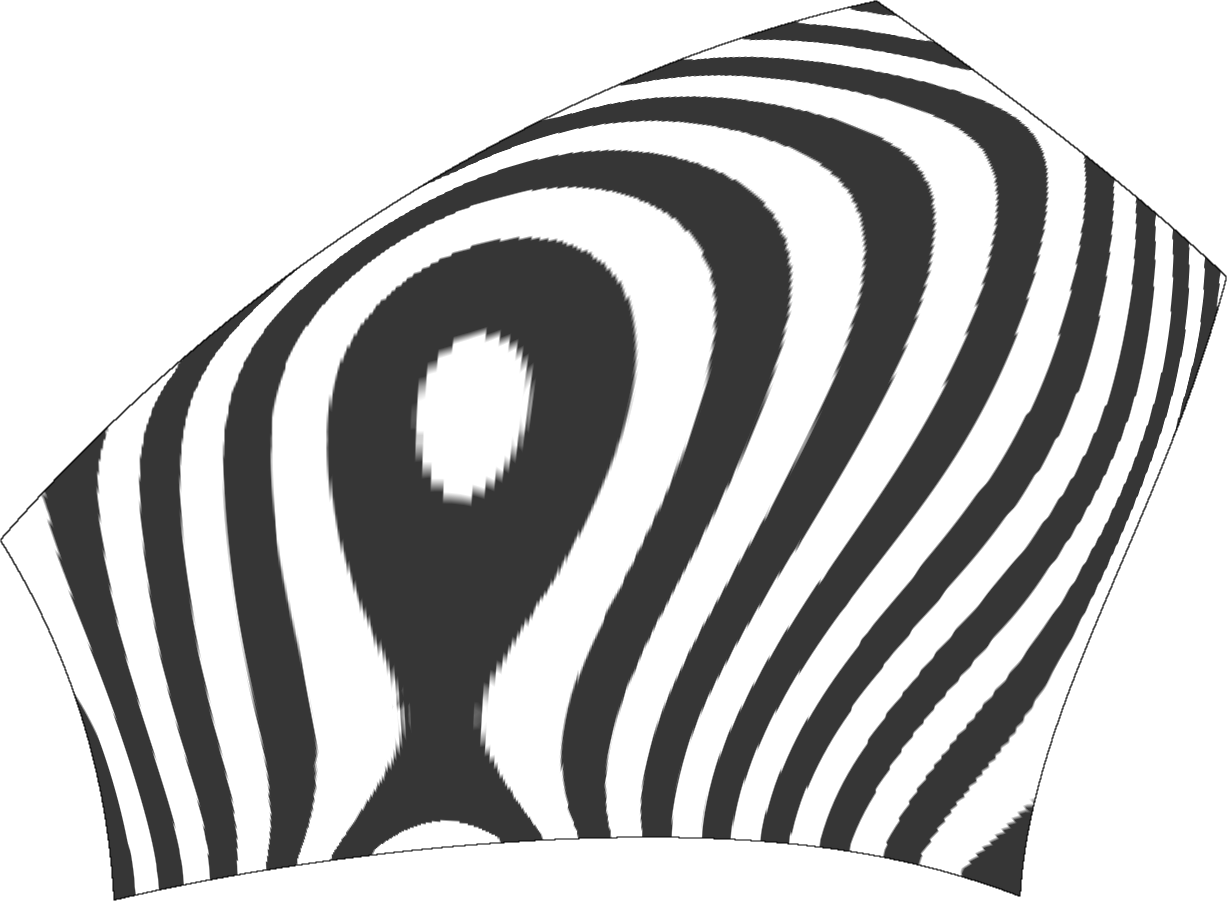}
\par\end{centering}
\caption{\label{fig:Isophotes}Isophotes of the surface shown in Figure~\ref{fig:Interior-control-points}.}
\end{figure}

Surface quality of S-patches generated by the above procedure are
generally very good, an example is shown in Figure~\ref{fig:Isophotes}.
The 5-sided quintic Bézier ribbon comprises 40 control points, while
the degree-elevated S-patch has 495 points (135 in boundary panels,
360 in the interior).

\begin{figure}
\begin{centering}
\hfill{}\includegraphics[width=0.46\textwidth]{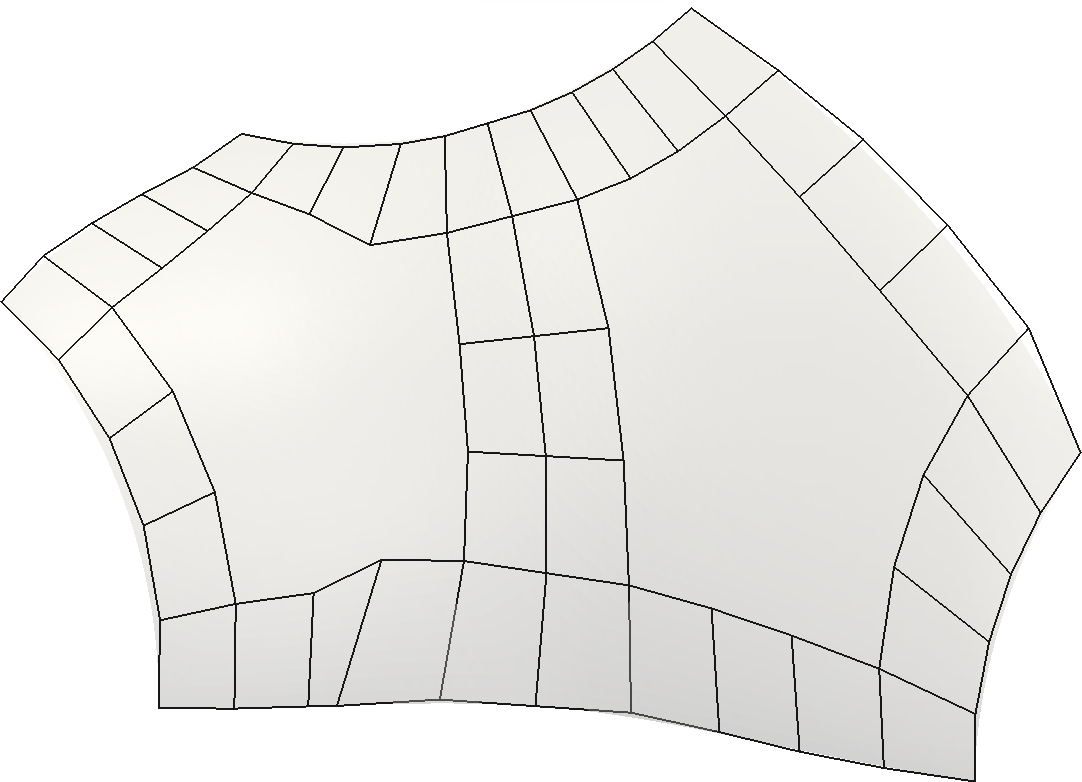}\hfill{}\includegraphics[width=0.47\textwidth]{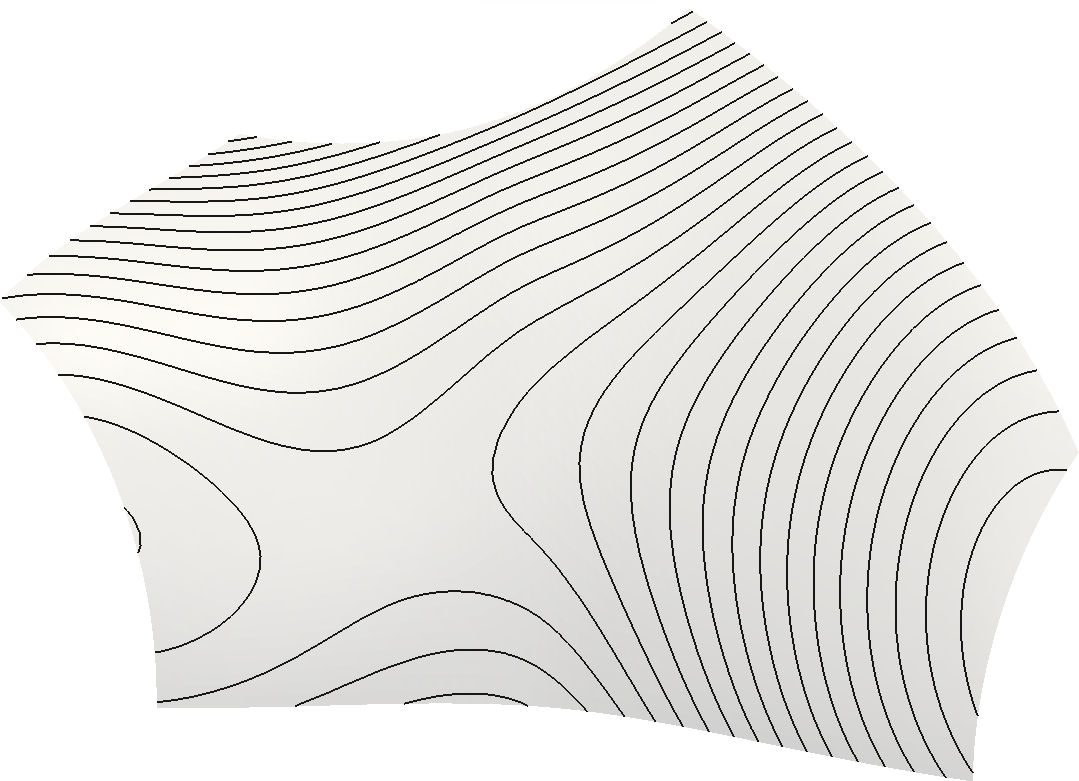}\hfill{}
\par\end{centering}
\caption{\label{fig:connections}S-patches connected with $G^{1}$ continuity
(left: ribbons, right: contours).}
\end{figure}

$G^{1}$ connection between adjacent surfaces is illustrated in Figure~\ref{fig:connections}.

\section*{Conclusion and future work}

We have seen that S-patches can be useful as surfaces filling boundary
loops in curve networks. An explicit construction was shown that achieves
tangential continuity with adjacent surfaces by fixing the boundary
panels of the patch. For the placement of the unconstrained control
points, a new algorithm was presented that can be used with surfaces
of any degree.

There are several interesting questions remaining, such as the effect
of the masks on the surface (does it minimize some functional?), or
a generalization to curvature continuous connections.

\section*{Acknowledgements}

This work was supported by the Hungarian Scientific Research Fund
(OTKA, No.\ 124727). The author thanks Tamás Várady for his valuable
comments.

\bibliographystyle{plain}
\bibliography{cikkek,sajat}

\end{document}